\documentclass[aps,twocolumn,raggedbottom,prl,showpacs,nobalancelastpage,amssymb,groupedaddress]{revtex4}
\usepackage{graphicx}
\usepackage{amsmath}
\begin{document}
\title{Entanglement spectroscopy of a driven solid-state qubit and its detector}
\author{M.\ C.\  Goorden$^1$, M.\  Thorwart$^2$, and M.\ Grifoni$^3$}
\affiliation{$^1$Instituut-Lorentz, Universiteit Leiden, P.O. Box 9506,
2300 RA Leiden, The Netherlands\\
$^2$Institut f\"{u}r Theoretische Physik IV,
Heinrich-Heine-Universit\"{a}t D\"{u}sseldorf, 40225
D\"{u}sseldorf, Germany\\$^3$Institut f\"{u}r Theoretische Physik,
Universit\"at Regensburg, 93035 Regensburg, Germany}
\date{\today}
\begin{abstract}
We study the asymptotic dynamics of a driven quantum two level
system coupled via a quantum detector to the environment. We find
multi-photon resonances which are due to the entanglement of the
qubit and the detector. Different regimes are studied by employing
a perturbative Floquet-Born-Markov approach for the qubit+detector
system, as well as non-perturbative real-time path integral
schemes for the driven spin-boson system. We find analytical
results for the resonances, including the red and the blue
sidebands. They agree well with those of exact ab-initio
calculations.
\end{abstract}
 \maketitle

A prominent physical model to study dissipative and decoherence
effects in quantum mechanics is the spin-boson model \cite{Wei}.
Currently, we witness its revival since it allows a
quantitative description of solid-state quantum bits (qubits)
\cite{Makhlin01}. A more realistic description requires  the inclusion
 of the external control fields  as well
as the detector. In the spin-boson
model, the environment is characterized  by a spectral
density $J(\omega)$. In its widest used  form, $J(\omega)$
is proportional to the frequency $\omega$  mimicking the effects
of an Ohmic electromagnetic environment. However,
 if the environment is
formed by a quantum detector which itself is damped by
Ohmic fluctuations, the simple  Ohmic description might become
inappropriate.
As an example, we focus on  a superconducting ring with three
Josephson junctions (so termed flux-qubit). It is read out by a
dc-SQUID \cite{Wal00,Chi03,Patrice04} whose plasma resonance at $\Omega_p$
gives rise to an effective spectral density $J_{\rm eff}(\omega)$
for the qubit with a peak at $\Omega_p$
\cite{Tian02}, cf.\  Eq.\  (\ref{jeff}) below. Recently, the
coherent coupling of a single photon mode and a superconducting charge
qubit has also been studied \cite{Wallr04}. Until now, the effects
of such a structured spectral density on decoherence {\em and\/}
in presence of a resonant control field have only been studied in
\cite{Thorwart00,Smi03} within a perturbative approach in $J_{\rm
eff}$. It was shown in \cite{Tho03,Kle03} for the
static case that a perturbative approach
breaks down for strong qubit-detector coupling, and
when the qubit and detector
 frequencies are comparable.

 In the presence of microwaves, multi-photon resonances are
expected to occur when the frequency of the ac-field, or integer
 multiples of it, match characteristic energy scales of the
 system \cite{Gri98}.  Such multiphoton resonances
 can be experimentally detected in an ac-driven flux qubit
 by measuring  the
  asymptotic occupation probabilities of the qubit, as the
  dc-field is varied  \cite{Wal00,Saito04}.
 %
 These ``conventional'' resonances, 
 which have also been theoretically investigated in \cite{Goo03},
 could be explained in terms  of intrinsic
 transitions in a driven
  spin-boson system with an unstructured environment.

  In this Letter, we investigate the asymptotic dynamics of a quantum
 two state system (TSS)
  with a structured
 environment, simultaneously driven by dc- and ac-fields.
 We show that a strong coupling between qubit and
  detector, together with the presence of a control field,
yields a non trivial dynamics involving additional  resonances
 in the entangled qubit+detector system.
 Our results are in agreement with recent experimental findings where such
``unconventional'' multi-photon transitions have been observed
\cite{Patrice04}.
\begin{figure}
\includegraphics[width=73mm,keepaspectratio=true]{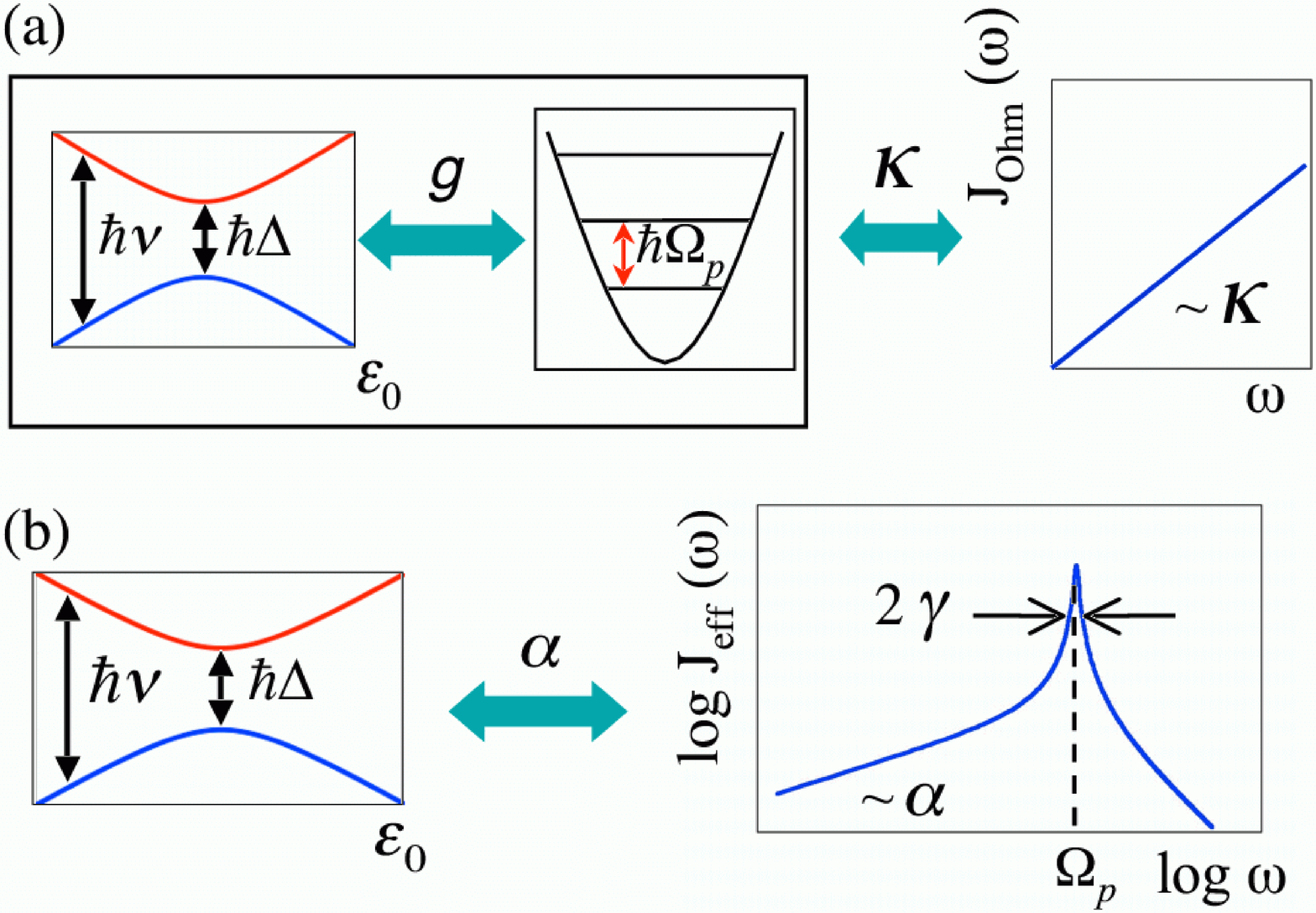}
\caption{Schematic picture of the models we use. In ($a$) the
system is a two-level-system (TSS) coupled to a harmonic
oscillator with the latter coupled to an Ohmic environment with
spectral density $J_{\rm Ohm}(\omega)$. In $(b)$, the TSS is
coupled to an environment with  peaked spectral density $J_{\rm
eff}(\omega)$.} \label{method}
\end{figure}
 We evaluate  the TSS dynamics in two completely equivalent
models,
cf.\  Fig.\  1. In  model $(a)$, the TSS is coupled to its detector
being represented as a single harmonic oscillator (HO)
 mode with frequency $\Omega_{p}$ with interaction strength $g$.
 The HO  itself interacts with a set of
 harmonic oscillators,
 cf.\  Fig.\ 1a. The corresponding Hamiltonian is
 $H_{QOB} (t) =  H_{QO} (t) + H_{OB}$, where
\begin{eqnarray}
H_{QO} (t)&=& -\frac{\hbar \Delta}{2} \sigma_x -
\frac{\hbar\varepsilon(t)}{2} \sigma_z + \hbar g \sigma_z X  +
\hbar \Omega_{p} B^{\dagger} B\nonumber \\
H_{OB}  & = &  X \sum_k \hbar \nu_k
(b_k^{\dagger} + b_k) + \sum_k \hbar \omega_k b_k^{\dagger} b_k  \, \nonumber\\
& & +  X^2 \sum_k \hbar \frac{\nu_k^2}{\omega_k} \, .
 \label{hamtlsosc}
\end{eqnarray}
Here, $\sigma_i$ are Pauli matrices, $\hbar\Delta$ is the
tunnel splitting, and  $\varepsilon(t)=\varepsilon_0 +
s\cos(\Omega t)$ describes the time-dependent
 driving with
the static bias $\varepsilon_0$. For $s=0$, the level splitting of
the isolated TSS is
$\hbar\nu=\hbar\sqrt{\varepsilon_0^2+\Delta^2}$.
Moreover,  $B$ is 
the annihilation 
operator of the localized HO mode, $X=B^{\dagger}+B$, while $b_k$
denote the bath mode operators. The
spectral density of the continuous bath modes is Ohmic with
dimensionless damping strength $\kappa$,  i.e.,
\begin{equation}
J_{\rm Ohm}(\omega) = \sum_k \nu_k^2 \delta(\omega-\omega_k) =
\kappa\omega \frac{\omega_D^2}{\omega^2+\omega_D^2} \, ,
\label{johm}
\end{equation}
where we have introduced a high-frequency cut-off at $\omega_D$.
In this approach, we shall consider
 the combined TSS + HO as the central
quantum system.

In the second model $(b)$, we exploit  the exact one-to-one
mapping \cite{Garg85} of the Hamiltonian (\ref{hamtlsosc}) onto that of
a driven spin-boson Hamiltonian \cite{Gri98}
\begin{eqnarray}
{H_{SB}}(t)&=&-\frac{\hbar\Delta}{2} \sigma_x
-\frac{\hbar\varepsilon(t)}{2} \sigma_z \nonumber\\
&+& \frac{1}{2}\sigma_z \hbar \sum_k \tilde{\lambda}_k
(\tilde{b}_k^{\dagger} +\tilde{b}_k) + \sum_k \hbar
\tilde{\omega}_k \tilde{b}_k^{\dagger} \tilde{b}_k\, ,
\label{hamtot}
\end{eqnarray}
where $\tilde{b}_k$ 
is the annihilation 
operator of the $k-$th bath mode with frequency
$\tilde{\omega}_k$.
Following \cite{Tian02}, the spectral density
 has a Lorentzian peak of width
 $\gamma=2\pi\kappa\Omega_p$ at the
characteristic detector frequency $\Omega_{p}$. It
behaves Ohmically at low frequencies
with the dimensionless coupling strength $\alpha
= \lim_{\omega \rightarrow 0} J_{\rm eff}(\omega)/2\omega$ and
reads
\begin{eqnarray}
\!\! J_{\rm eff}(\omega)=\sum_k \tilde{\lambda}_k^2
\delta(\omega-\tilde{\omega}_k)=  \frac{2 \alpha \omega
\Omega_{p}^4}{(\Omega_{p}^2-\omega^2)^2+(\gamma\omega)^2}.
%
%
\label{jeff}
\end{eqnarray}
The relation between $g$ and $\alpha$ follows as
$g=\Omega_{p} \sqrt{\alpha/8 \kappa}$.
In this model, we associate the detector as part
of the qubit environment.

The qubit dynamics is described by the reduced density operator
$\rho(t)$ obtained by tracing out all environmental degrees of freedom.
We study the population difference
$P(t):=\langle\sigma_z\rangle(t)=tr(\rho (t) \sigma_z)$ in
 the asymptotic
limit, i.e.,   $P_\infty=\lim_{t\rightarrow\infty}\langle
P(t)\rangle_\Omega$, where the averaging is over one period of the
ac-field.
%
%
\begin{figure}
\includegraphics[width=8.5cm]{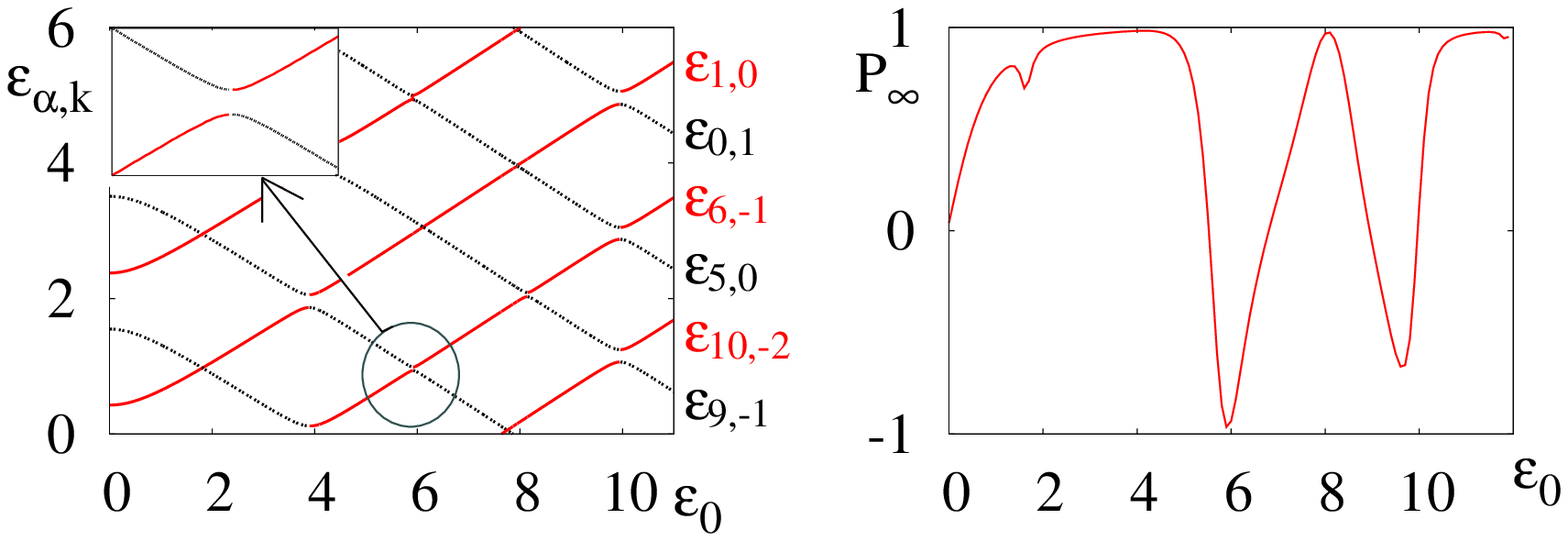}
\caption{Left: Quasi-energy spectrum $\varepsilon_{\alpha,k}$ of
the driven TSS+HO system vs dc-bias  $\varepsilon_0$ (in units of
$\Delta$). The quasi-energies are defined up to an integer
multiple of $\hbar \Omega$, i.e.,
$\varepsilon_{\alpha,k}=\varepsilon_{\alpha} +k\hbar \Omega$.
Inset: Zoom of an anti-crossing. Right: $P_{\infty}$ exhibits
sidebands corresponding to quasi-energy level anti-crossings.
Parameters are $\Omega=10\Delta, s=4\Delta, g=0.4\Delta,
\Omega_p=4\Delta, \kappa=0.014$ and $k_B T=0.1\hbar\Delta$.}
 \label{spectrum}
\end{figure}

%
{\em Case of weak damping and low temperatures}.
For $\kappa \ll 1$ and $k_BT \lesssim \hbar\Delta$,
it is convenient to use model ($a$). The equations of motion for
the TSS+HO reduced density
 matrix are most conveniently derived in the Floquet basis \cite{Blu89}.
The Floquet states $|\phi_{\alpha}(t)\rangle=\sum_n|\phi_{\alpha}^{(n)}
\rangle\exp(in\Omega t)$
corresponding to a periodic Hamiltonian $H(t)$ can be obtained from the eigenvalue equation
${\cal{H}}|\phi_{\alpha}(t)\rangle=\varepsilon_{\alpha}|\phi_{\alpha}
(t)\rangle$, with the
Floquet Hamiltonian ${\cal{H}}=H(t)-i\hbar\frac{\partial}{\partial t}$.
%
Upon including dissipative effects to lowest order in $\kappa$, a
Floquet-Born-Markov master equation is obtained \cite{Gri98,Kohler97}. We
average the $2\pi/\Omega$-periodic coefficients 
over one period of the driving, assuming that dissipative
effects are relevant on much larger timescales.
In the Floquet basis, this yields
 equations of motions for
$\rho_{\alpha\beta}(t)=\langle\phi_{\alpha}(t)|\rho(t)|
\phi_{\beta}(t)\rangle$
of the form
\begin{eqnarray}
\dot{\varrho}_{\alpha\beta}(t)
=-\frac{i}{\hbar}(\varepsilon_{\alpha}-\varepsilon_{\beta})
\varrho_{\alpha\beta}(t)+
\sum_{\alpha'\beta'}L_{\alpha\beta,
\alpha'\beta'}\varrho_{\alpha'\beta'}(t),
\label{rhoeq}
\end{eqnarray}
with the dissipative transition rates
\begin{eqnarray}
L_{\alpha\beta,
\alpha'\beta'}&=&\sum_n(N_{\alpha\alpha',n}
+N_{\beta\beta',n})X_{\alpha\alpha',n}
X_{\beta'\beta,-n}\nonumber\\
&-&\delta_{\beta\beta'}\sum_{\beta'',n}N_{\beta''
\alpha',n}X_{\alpha\beta'',-n}X_{
\beta''\alpha',n}\nonumber\\
&-&\delta_{\alpha\alpha'}\sum_{\alpha'',n}N_{
\alpha''\beta',n}X_{\beta'\alpha'',-n}X_{\alpha''
\beta,n}.
\label{rates}
\end{eqnarray}
Here $X_{\alpha\beta,n}=\sum_k\langle
\phi^{(k)}_{\alpha}|X|\phi^{(k+n)}_{\beta}\rangle$, and
$N_{\alpha\beta,n}=N(\varepsilon_{\alpha}-
\varepsilon_{\beta}+n\hbar\Omega)$
with
$N(\varepsilon)=\frac{\kappa\varepsilon}{2\hbar}(\coth{(\frac{\varepsilon}{2
k_B T})}-1)$ (assuming $\omega_D\rightarrow\infty$).

Following  Ref.\  \cite{Shi65} we write the Floquet Hamiltonian ${\cal{H}}_{QO}$
in the basis
$|a,n\rangle$, with $|a\rangle=|g/e,m\rangle$, $g/e$
being the ground/excited state of the qubit, $m$ the oscillator
state, and $n$ the Fourier index. In this basis, ${\cal
H}_{QO}$ has diagonal elements ${\cal{H}}_{a n,a
n}=\hbar[\mp\nu/2+m\Omega_p+n\Omega]$, and off-diagonal elements
$V_{a{n},b{k}}=\langle a|\delta_{n,k}\hbar g X
\sigma_z+(\delta_{n,k+1}+\delta_{n+1,k})\frac{\hbar
s}{4}\sigma_z|b \rangle$.
%
%
The quasi-energy spectrum of ${\cal H}_{QO}$
is shown in Fig.\ 2 as a function of the
bias $\varepsilon_0$.
We find avoided level crossings when $E_{a n, b m}:={\cal{H}}_{a
n, a  n}-{\cal{H}}_{ b m, b m}=0+O(V^2)$, i.e., ($m \ge 0, -\infty
< n < +\infty $)
\begin{equation}
 \nu=n\Omega\pm m\Omega_{p}+O(V^2),
 \quad  n\Omega=m\Omega_{p} +O(V^2) \, .
\end{equation}
%
 Associated to the avoided crossings  are
resonant peaks/dips of $P_\infty$, see Fig.\ 2. The resonances at
$\nu=n\Omega\pm m\Omega_{p}$ are known as red/blue   sidebands
\cite{Cohbook}.

\begin{figure}
\includegraphics[width=9cm]{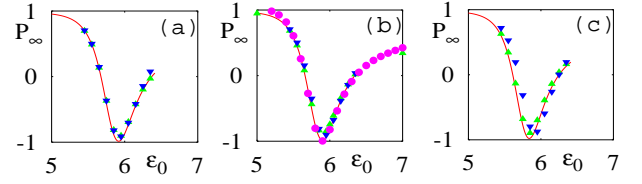}
\caption{First blue sideband $P_\infty$ vs $\varepsilon_0$ (in
units of $\Delta$) at $\nu=\Omega-\Omega_p$.
 The solid lines are the analytical prediction (\ref{analytical}) for
 $(a)$ $g=0.05\Delta$, $(b)$ $g=0.2\Delta$, $(c)$ $g=0.4\Delta$.
 The triangles are the results of a Floquet-Bloch-Redfield simulation, cf.  Eq.
 (\ref{rhoeq}),
 with one (upward triangles) and two (downward triangles) HO levels taken into account.
 The circles in ($b$) are the results from a QUAPI simulation with six HO levels.
 We choose $s=2\Delta$,
 $\Omega=10\Delta$, $\Omega_p=4\Delta$, $\kappa=0.014$, $k_BT=0.1\hbar\Delta$. } \label{floquet}
\end{figure}

In the following, we derive an analytical expression for the first
blue sideband at $\nu\approx\Omega-\Omega_{p}$. Other resonances
can be evaluated in the same way.
 We  include only one
HO level ($m=0,1$) which is appropriate
because we investigate a resonance between $|g,0\rangle$ and
$|e,1\rangle$ with $g/\Omega_p\ll 1$.  We consider $V_{a
n, b k}$ as a perturbation, and use the method of Ref.\
\cite{Cohbook,Sha80} to obtain an effective Hamiltonian
${\cal{H}}_{\rm{eff}}=e^{iS}{\cal{H}}_{QO}e^{-iS}$, with
\begin{eqnarray}
iS_{a n,b m}&=&\left[\sum_{c, k}\frac{V_{a
n,c k}V_{c k,b m}}{2E_{b m,a
n}}\left(\frac{1}{E_{c k,a n}}
+\frac{1}{E_{c k,b m}}\right)\right.\nonumber\\
&+&\left.\frac{V_{a n,b m}}{E_{a n,b m}}\right]\mbox{
for $|E_{a n,b m}|\neq|\nu+\Omega_p-\Omega| $},
\end{eqnarray}
and $iS_{a n,b m}=0$ for $|E_{a n,b m}|=|\nu+\Omega_p-\Omega|$.
The block-diagonal ${\cal{H}}_{\rm{eff}}$  has the same
eigenvalues as ${\cal{H}}_{QO}$ with quasi-degenerate eigenvalues
$\varepsilon_{1,2}$ in one block.  With
$c_{1/3}=\frac{g^2}{\nu^2}(\frac{-\varepsilon_0^2}{\Omega_{p}}\mp\frac{\Delta^2}{\nu\pm\Omega_{p}})\mp
\frac{\Delta^2s^2}{8(\nu^2-\Omega^2)\nu}$  and
$\delta=\nu-\Omega+\Omega_{p}-2c_1$ the  quasi-energies {\em up to
second order in $V$} read
\begin{eqnarray}
\varepsilon_{1/2}&=&
-\hbar\nu/2+\hbar\delta/2\,(1\mp\sqrt{1+\Delta_1^2/\delta^2})+\hbar c_1 \, , \nonumber\\
\varepsilon_3&=&-\hbar\nu/2+\hbar\Omega_{p}-\hbar c_3,
\qquad\hspace{-0.5cm} \varepsilon_4= \hbar\nu/2+\hbar c_3.
\end{eqnarray}
The eigenvalue splitting at the level crossing, $\delta=0$, is
\begin{equation}
\hbar\Delta_1=\frac{\hbar\Delta\varepsilon_0gs\left
[\Omega^2+\Omega_p^2+\nu(-\Omega+\Omega_p)\right]}
{4\nu(\Omega-\nu)\Omega\Omega_p(\nu+\Omega_p)}.
\end{equation}
The Floquet states are, with $\tan{\theta}=2|\Delta_1|/\delta$,
$B^+(x)=\cos(x)$ and $B^{-}(x)=\sin(x)$,
\begin{eqnarray}
|\phi_{1/2}\rangle&=&e^{-iS}[B^{\mp}{(\theta/2)}e^{-i\Omega t}|e,1\rangle\pm B^\pm{(\theta/2)}|g,0\rangle],\nonumber\\
|\phi_3\rangle&=&e^{-iS}|g,1\rangle , \qquad
|\phi_4\rangle=e^{-iS}|e,0\rangle. \label{Floquet}
\end{eqnarray}
With this, we can calculate the rates in Eq.\  (\ref{rates}) up to
second order in $V$. It holds,
$L_{1122}=L_{2211}=L_{3344}=L_{4433}=0$.
To find  the stationary state of Eq.\
(\ref{rhoeq}),  we assume that 
$\rho_{\alpha\beta}(\infty)=0$ for $\alpha\neq\beta$, except for
$\rho_{12}$ and $\rho_{21}$ (secular approximation). This is valid
 if
$\varepsilon_\alpha-\varepsilon_\beta\gg
L_{\alpha\beta,\alpha'\beta'}$, which is true for
non-quasi-degenerate eigenvalues because $\kappa\ll 1$. We find 
at resonance 
\begin{eqnarray}
P_{\infty}=-\frac{\varepsilon_0}{\nu}\tanh\left(\frac{\hbar\Omega_p}{2
k_B T}\right)+O\left(V^2\right), \label{pinffloquet}
\end{eqnarray}
implying a complete inversion of population at low temperatures.
%
%
%
%
Far enough off-resonance, we can  
assume that
$\rho_{12}(\infty)=\rho_{21}(\infty)=0$ and $\sin{(\theta/2)}\simeq \theta/2$.
We presume
$k_BT\ll\hbar\Omega_p,\hbar\Omega,\hbar\nu$ (which allows to
set $N(\hbar\Omega_p)=N(\hbar\Omega)=N(\hbar\nu)=0$) and find the major result
\begin{eqnarray}
P_{\infty}&=&\frac{\varepsilon_0}{\nu}\frac{L_{1144}-L_{4411}}{L_{1144}+L_{4411}}
+O\left(V^2\right)  \nonumber \\
& \simeq & \frac{\varepsilon_0}{\nu}\left(
1 - \frac{2\Delta_1^2 \nu^2 (\nu^2-\Omega_p^2)^2}
{\Delta_1^2 \nu^2 (\nu^2-\Omega_p^2)^2 + 4 \Delta^2 g^2 \Omega_p^2\delta^2}
\right)\!\!.\,
\label{analytical}
\end{eqnarray}
Because the oscillator can give its energy directly to the
environment,   the decay from $|e/g,1\rangle$ to $|e/g,0\rangle$
is much faster than the other processes and does not play a role
in Eq.\  (\ref{analytical}). Hence, $P_\infty$ is determined by the
ratio of two rates: $L_{4411}\sim\sin^2(\theta)\sim s^2g^2$ which
describes the timescale of driving induced transitions from
$|g,0\rangle$ to $|e,0\rangle$, and $L_{1144}\sim g^2$ for the
qubit decay from $|e,0\rangle$ to $|g,0\rangle$ via the
oscillator.  Since both scale as $g^2$ we
find for this particular resonance
that $P_{\infty}$ is {\em independent} of $g$. 
%
Fig.\  \ref{floquet} shows the results of Eq.\
(\ref{analytical}) and
different numerical results, including those of an ab-initio
real-time  QUAPI \cite{QUAPI} calculation.
%
A good agreement, even near resonance, is
found. A similar analysis yields
$P_\infty=\frac{\varepsilon_0}{\nu}\tanh({\frac{\hbar\Omega_p}{2k_B
T}})+O(V^2)$ for the first red sideband at $\nu=\Omega+\Omega_p$,
which is very close to thermal equilibrium for low $T$.
For $\nu=\Omega$ only the oscillator is
excited and
thermal equilibrium $P_\infty=\frac{\varepsilon_0}{\nu}\tanh({\frac{\hbar\nu}{2 k_B
T}})$ is recovered.

\begin{figure}
\includegraphics[width=5cm,angle=270]{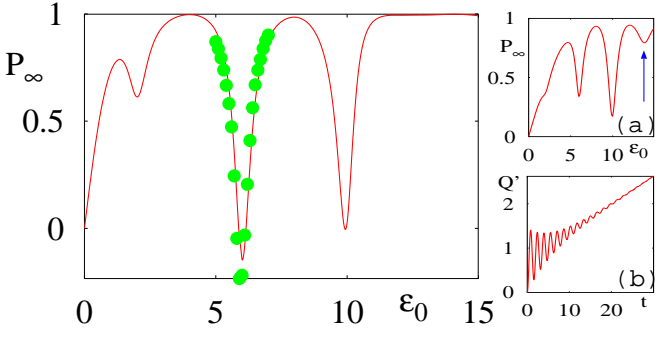}
\caption{ $P_\infty$ vs $\varepsilon_0$ (in units of $\Delta$).
The solid line is the NIBA prediction, while the circles are from
a QUAPI simulation with 6 HO levels ($g=3\Delta$, $s=4\Delta$,
$\Omega=10\Delta$, $\kappa=0.014$, $k_BT=0.5\hbar\Delta$,
$\Omega_p=4\Delta$). Inset ($a$): NIBA result for $k_B
T=2\hbar\Delta$. The arrows indicates the first red sideband at
$\nu=\Omega+\Omega_p$. Inset ($b$): $Q'(t)$ vs $t$ shows damped
oscillations. } \label{niba}
\end{figure}

{\em  Case of strong damping and/or high temperatures}.
In the complementary regime of large environmental
coupling and/or high temperatures it is convenient to  employ
model ($b$),
and is appropriate to treat the system dynamics within the
noninteracting-blip approximation (NIBA) \cite{Wei}. The NIBA  is
non-perturbative in the coupling  $\alpha$
 but perturbative in the tunneling splitting $\Delta$.
 Within the NIBA, and for large driving frequencies $\Omega > \Delta
 $, one finds $P_{\infty}={k_0^-}/{k_0^+}$ \cite{Gri98}, where
\begin{eqnarray}
 k^{\pm}_0=\Delta^2 \int_0^\infty d\tau
h^{\pm}(\tau)B^{\pm}(\varepsilon_0
\tau)J_{0}\left(\frac{2s}{\Omega}\sin{\frac{\Omega
\tau}{2}}\right). \label{kernels}
\end{eqnarray}
The influence of the dc- and ac-field is in the terms $B^{\pm}(x)$,
and in the Bessel function $J_0$,
respectively.
 Dissipative effects are
 captured by $h^{\pm}(t)=e^{-Q'(t)}B^{\pm}[Q''(t)]$, where
$Q'(t)$ and $Q''(t)$ are the real and imaginary parts of the bath
correlation function\cite{Gri98}. 
%
For the peaked  spectral density Eq.\  (\ref{jeff}) one finds
%
\begin{eqnarray}
Q'(t)&=& Q'_1(t)-e^{-\Gamma t}[Y_1\cos(\bar{\Omega}_{p}t)+Y_2\sin(\bar{\Omega}_{p}t)] \nonumber\\
 Q''(t)&=&A_1-e^{-\Gamma
t}[A_1\cos(\bar{\Omega}_pt)+A_2\sin(\bar{\Omega}_{p}t)] \, .
\end{eqnarray}
Here,  $\beta=\hbar/k_B T, \Gamma=\pi\kappa\Omega_{p}$,
$\bar{\Omega}_{p}^2=\Omega_{p}^2-\Gamma^2$ and
%
%
\begin{eqnarray}
Q'_1(t)&=&
Y_1+\pi\alpha\Omega_p^2 \left[\frac{\sinh(\beta\bar{\Omega}_p)
 t}{2C\bar{\Omega}_p}+\frac{\sin(\beta\Gamma) t}{2C\Gamma} \right.
 \nonumber \\
&&  \left. -\frac{4\Omega_{p}^2}{\beta}\sum_{n=1}^\infty
\frac{\frac{1}{\nu_n}[e^{-\nu_n
t}-1]+t}{(\Omega_p^2+\nu_n^2)^2-4\Gamma^2\nu_n^2} \right],
\label{kern}
\end{eqnarray}
where $\nu_n=2\pi n/\beta$. Moreover,
$C=\cosh(\beta\bar{\Omega}_{p})-\cos(\beta\Gamma)$, $CY_{1/2}=\mp
A_{2/1}\sinh\beta\bar{\Omega}_{p}-A_{1/2}\sin\beta\Gamma$,
$A_2=\alpha\pi(\Gamma^2-\bar{\Omega}_p^2)/2\Gamma\bar{\Omega}_p$,
$A_1=\pi\alpha$. So, $Q'$ and $Q''$
display damped oscillations (cf.\ Fig.\ \ref{niba}) not present for
a pure Ohmic spectrum. It is the interplay between these
oscillations and the driving field which induces the extra
resonances in $P_\infty$.
 In the regime $\Gamma/\Omega_p \ll 1$, the term
$\exp(-\Gamma t)$  in Eq.\  (\ref{kern}) varies slowly on
the time-scale of the oscillations. We expand $Q'$ and $Q''$ as well as the Bessel function $J_0$ entering (\ref{kernels}) using Bessel function identities and find the important result
\begin{equation}
k_0^{\pm}=\sum_{m=0}^{\infty}\sum_{n=-\infty}^\infty\Delta^2\int_0^\infty
dt e^{-Q_1'(t)}f^{\pm}_{mn}(t)\;,
\end{equation}
where $\varepsilon_{mn}=\varepsilon_0-m\bar\Omega_p-n\Omega$, and
\begin{eqnarray}
\label{expansion} &&f^{\pm}_{mn}(t)={{\rm Re} \atop {\rm
Im}}\left[{c^{\pm}_{mn}(t)\cos(\varepsilon_{mn}t)
\pm c^{\mp}_{mn}(t)\sin(\varepsilon_{mn}t)}\right],\nonumber\\
&&c^\pm_{mn}=J_n^2\left(\frac{s}{\Omega}\right)J_m(e^{-\Gamma t}\omega_1)
B^{\pm}(m\phi)(-i)^m e^{-iA_1} \, .
\end{eqnarray}
Here is $\omega_1=\sqrt{(A_1-iY_1)^2+(A_2-iY_2)^2}$, and
$\tan\phi=-({A_2-iY_2})/({A_1-iY_1})$. Thus, from Eq.\
(\ref{expansion}) we expect resonances when $\varepsilon_{nm}=0$.
Without driving we always find that around
$\varepsilon_0=m\bar{\Omega}_p$ it holds
$P_\infty\approx\tanh ({m\beta{\Omega}_p/2)}$,
since
$\lim_{\Gamma/\Omega_{p}\rightarrow
0}\tan(m\phi)=i\tanh(m\beta\Omega_{p}/2)$
(for not too large $T$, i.e.,  $\cos(\beta\Gamma)\ll\cosh(\beta\Omega_p)$).
Hence, $P_\infty$ acquires its NIBA thermal equilibrium value,
 and driving is needed to
see resonances. For ``conventional'' resonances at
$\varepsilon_0=n\Omega$ we find $P_\infty\approx 0$, as predicted
for unstructured  environments \cite{Har00,Goo03}. Finally, for
$\varepsilon_0=n\Omega\pm m\bar{\Omega}_{p}$, we recover
$P_\infty\approx\pm\tanh(m \beta\Omega_{p}/2)$, as also was found
within the Floquet-Born-Markov approach, cf.\ (\ref{pinffloquet}).
 Results of a numerical evaluation of
$P_\infty$ are shown in Fig.\ 4, using the NIBA result
(\ref{expansion}), as well as the exact ab-initio real-time
 QUAPI method \cite{QUAPI}. In the numerical
evaluation, we could not reach the parameter regime
$\Gamma/\Omega_p\ll 1$, but still  clear resonance dips are
observed at $\varepsilon_0=\Omega$,
$\varepsilon_0=\Omega-\Omega_p$ and
$\varepsilon_0=\Omega-2\Omega_p$. For $k_B T\sim \hbar \Omega_p$,
we also find the first red sideband at
$\varepsilon_0=\Omega+\Omega_p$, see inset.

In conclusion we evaluated the asymptotic population of a driven TSS in
a structured environment.  
We have derived analytic expressions for the shape of the
resonances for both weak and strong damping. We show that the
coupling of the
 TSS and the detector is revealed in the occurrence of characteristic
 multi-photon resonances, also reported in recent experiments \cite{Patrice04}, 
  in the asymptotic population of the TSS. A complete population
  inversion is predicted for the blue-sidebands transitions, while
  values close to equilibrium are found for the red-sidebands.

 Support by the Dutch NWO/FOM,
 and the Universit\"atsstiftung Hans Vielberth, and discussions with P.\ Bertet, I. Chiorescu and H.\
Mooij are acknowledged. 

\end{document}